\documentclass[pre,twocolumn]{revtex4}
\usepackage{epsfig}
\usepackage{bm}
\usepackage{amssymb}
\usepackage{amsmath}

\begin{document}

\title{Chaos, randomization, and turbulence in particle-laden flows}

\author{A. Bershadskii}

\affiliation{
ICAR, P.O. Box 31155, Jerusalem 91000, Israel
}

\begin{abstract}

  The randomization effect of the two-way (particle-flow) interaction has been studied and quantified using the notion of distributed chaos and the results of numerical simulations and laboratory measurements. It is shown, in particular, that an increase of such parameters as the particle volume fraction, particle mass loading, and Stokes number results generally in stronger randomization of the particle-laden flows.  An important role of spontaneous breaking of the local reflectional symmetry in the randomization of the particle-laden flows has been also analyzed using relevant dynamical invariants.

\end{abstract}

\maketitle

\section{Introduction}

    A general classification of the non-laminar regimes in fluid dynamics can be based on the concept of smoothness. Smooth dynamics is characterized by  the {\it stretched} exponential spectra
$$
E(k) \propto \exp-(k/k_{\beta})^{\beta}.  \eqno{(1)}
$$     
 where $1 \geq \beta > 0$ and $k$ is wavenumer. The particular value $\beta =1$ - the exponential spectrum:
$$ 
E(k) \propto \exp(-k/k_c),  \eqno{(2)}
$$ 
is usually associated with deterministic chaos (see, for instance, Refs. \cite{fm}-\cite{kds}). 

  For $1 > \beta$ the chaotic-like dynamics is still smooth but not deterministic (the distributed chaos). The non-smooth dynamics is characterized by the power-law (scaling) spectra. \\
  
  In the distributed chaos the value of the parameter $\beta$ can be used as a measure of randomization: a lesser value of $\beta$ (i.e. it is further from the deterministic value $\beta =1$) corresponds to stronger randomization of the flow. \\ 
  
\begin{figure} \vspace{-0.3cm}\centering 
\epsfig{width=.45\textwidth,file=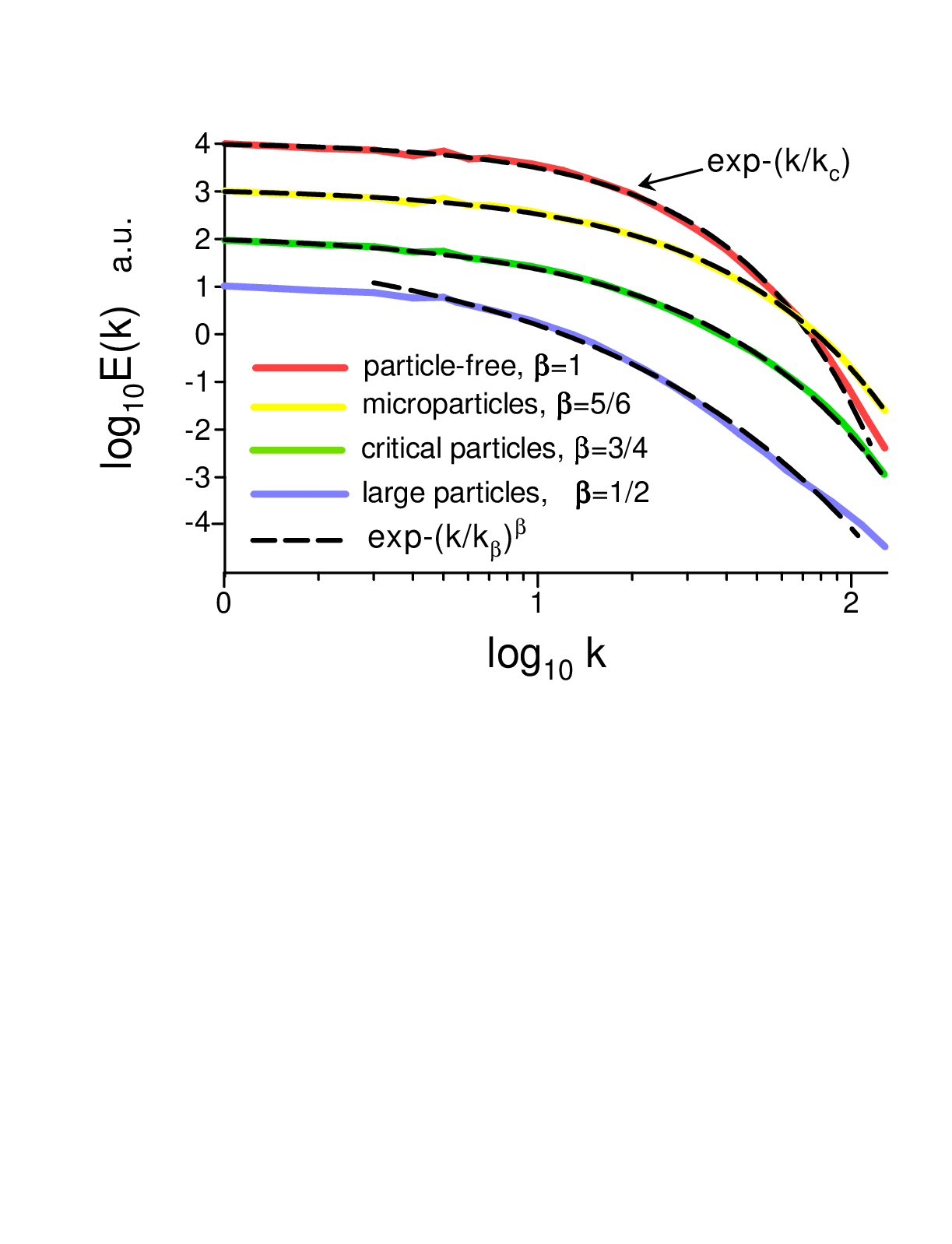} \vspace{-4.85cm}
\caption{Kinetic energy spectra computed in the isotropic decay at $t = 5$. The spectra are vertically shifted for clarity.} 
\end{figure}
\begin{figure} \vspace{-0.4cm}\centering 
\epsfig{width=.45\textwidth,file=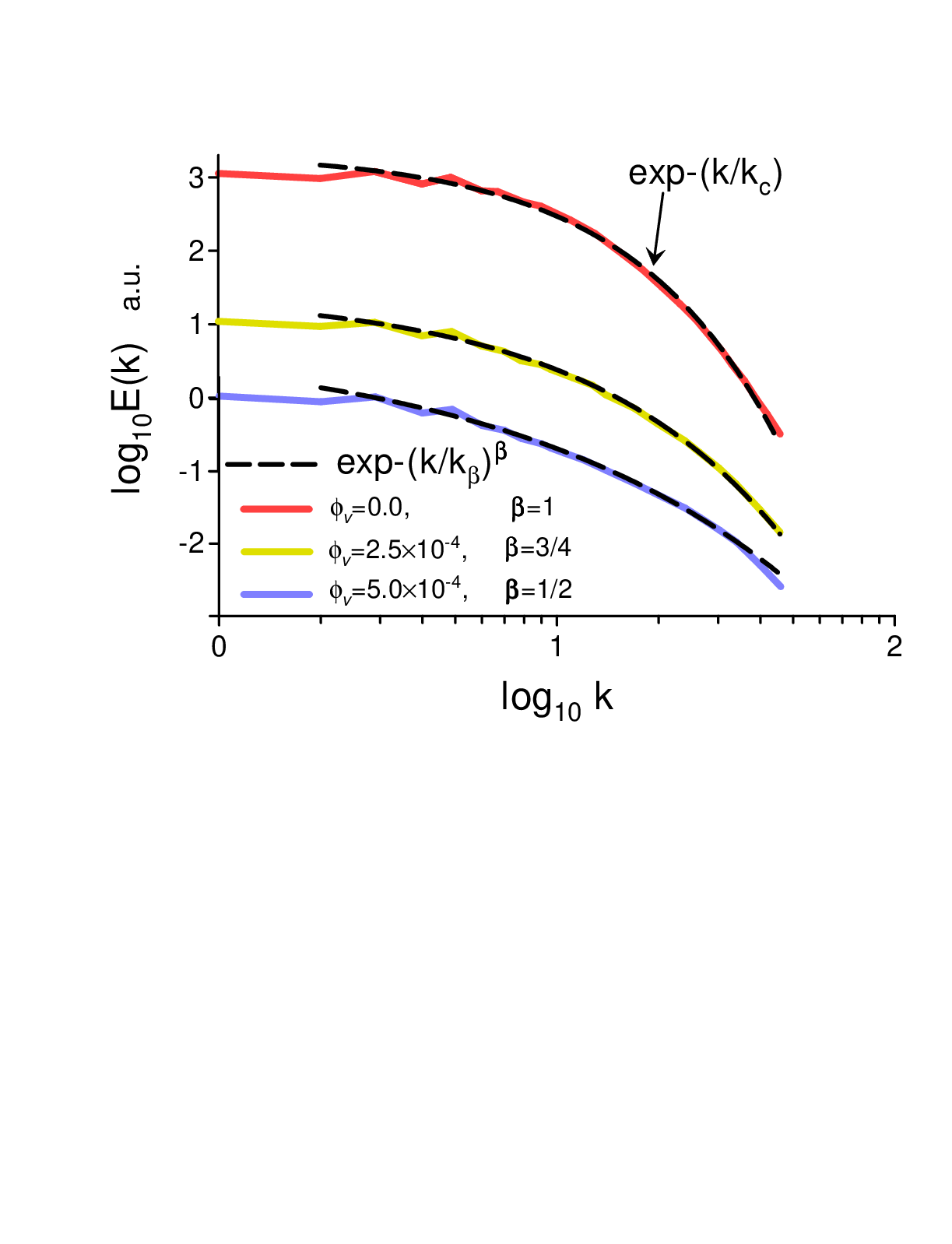} \vspace{-4.55cm}
\caption{Kinetic energy spectra computed in the isotropic decay at $t = 4$ for different values of the particle volume fraction $\phi_v$ . The spectra are vertically shifted for clarity.} 
\end{figure}

  In the particle-laden flows the particles can be an important additional source of the fluid dynamics randomization (whereas the increase in the Reynolds number is still the main source of randomization). Figure 1, for instance, shows the influence of the different types of particles on the fluid's kinetic energy spectrum. The spectral data were taken from Fig. 3 of the Ref. \cite{fe}. In this paper results of direct numerical simulation (DNS) of the decay of incompressible isotropic flow (with and without inertial particles) were reported (we will return to the description of this DNS in more detail below). Now one can see that for the unladen (particle-free) case the kinetic energy spectrum is exponential (it is indicated by the dashed curve), i.e. corresponds to the deterministic chaos (the Taylor-scale Reynolds number  $Re_{\lambda} = 31$ for the time of decay when this spectrum was computed). For the particle-laden cases, the spectra can be well fitted by the stretched exponentials Eq. (1) (the dashed curves), i.e. these cases correspond to the (smooth) distributed chaos. The values of the parameter $\beta$ are different for the different types of particles and a clear trend for randomization can be observed. 
  
  The analogous trend can be observed in Fig. 2 where results of an analogous DNS but with different values of the particle volume fraction $\phi_v$ have been shown (the spectral data were taken from Fig. 10 of Ref. \cite{et}). \\
  
   In the next sections we will relate the values of  $\beta$ observed in the Figs. 1 and 2 with invariants of the fluid dynamics and will show an important role of the spontaneous breaking of the local reflectional symmetry in the randomization of the particle-laden flows.

 \section{The fluid dynamics invariants and distributed chaos}  
 
When the characteristic scale $k_c$ in the spectrum Eq. (2) randomly fluctuates one should use an ensemble averaging to obtain the average spectrum 
$$
E(k) \propto \int_0^{\infty} P(k_c) \exp -(k/k_c)dk_c \eqno{(3)}
$$    
If the randomized dynamics is still smooth, then $E(k)$ has the stretched exponential form Eq. (1). Comparing Eq. (1) and Eq. (3) one can find asymptote of  the probability distribution $P(k_c)$ for large $k_c$ \cite{jon}
$$
P(k_c) \propto k_c^{-1 + \beta/[2(1-\beta)]}~\exp(-\gamma k_c^{\beta/(1-\beta)}), \eqno{(4)}
$$     
here $\gamma$ is a constant.\\

 The ideal (nondissipative) fluid dynamics has two fundamental invariants: energy and helicity. The dissipative (Navier-Stokes) dynamics also has two fundamental invariants: Birkhoff-Saffman integral \cite{bir},\cite{saf},\cite{dav}
$$   
I_{BS} = \int  \langle {\bf u} ({\bf x},t) \cdot  {\bf u} ({\bf x} + {\bf r},t) \rangle d{\bf r},  \eqno{(5)}
$$   
and Loitsyanskii integral \cite{dav} ,\cite{my}
$$
I_L =  \int r^2 \langle {\bf u} ({\bf x},t) \cdot  {\bf u} ({\bf x} + {\bf r},t) \rangle d{\bf r}  \eqno{(6)}
$$  
where $<...>$ denotes a global (ensemble or spatial) average, ${\bf u} ({\bf x},t)$ is the velocity field.   

  Due to the Noether’s theorem the conservation of the Birkhoff-Saffman and Loitsyanskii integrals by the Navier-Stokes dynamics is a consequence of the space homogeneity and isotropy respectively.\\
  
   Let us denote an invariant as $I$ and relate the characteristic velocity $u_c$ and the characteristic scale $k_c$ using dimensional considerations
 $$
 u_c \propto I^{\delta} k_c^{\alpha} \eqno{(7)}
 $$ 
where $\delta$ and $\alpha$ are some parameters that can be found from the dimensional considerations. \\

  For the normally distributed characteristic velocity $u_c$ \cite{my} the distribution $P(k_c)$ can be readily found from the relationship Eq. (7).  Comparing this distribution with the asymptote Eq. (4) one obtains a relationship between the exponents $\alpha$ and $\beta$
$$
\beta = \frac{2\alpha}{1+2\alpha}  \eqno{(8)}
$$  

  For instance, using the Birkhoff-Saffman integral Eq. (5) as invariant $I$ in Eq. (7) we obtain $\delta =1/2$ and $\alpha =3/2$, and consequently from the Eq. (8)
$$
E(k) \propto \exp-(k/k_{\beta})^{3/4}  \eqno{(9)}
$$  
For the flow dominated by the Loitsyanskii integral Eq. (6) $\delta =1/2$ and $\alpha = 5/2$, and consequently from the Eq. (8)
$$
E(k) \propto \exp-(k/k_{\beta})^{5/6}.  \eqno{(10)}
$$ 

  One can recognize these spectra in the Figs. 1 and 2.\\
  
  Let us now discuss the DNS reported in the Refs. \cite{fe},\cite{et} in more detail. 
  
  The dispersed solid particles (80 million) were injected in the freely decaying isotropic and homogeneous flow described by the Navier-Stokes equation   
$$
 \frac{\partial {\bf u}}{\partial t} = - ({\bf u} \cdot \nabla) {\bf u} 
    -\frac{1}{\rho} \nabla {p}  + \nu \nabla^2  {\bf u} +{\bf F}_p  \eqno{(11)}
$$  
for an incompressible fluid $ \nabla \cdot {\bf u} = 0$. The diameter of particles was smaller than the Kolmogorov scale. \\

\begin{figure} \vspace{-0.8cm}\centering
\epsfig{width=.44\textwidth,file=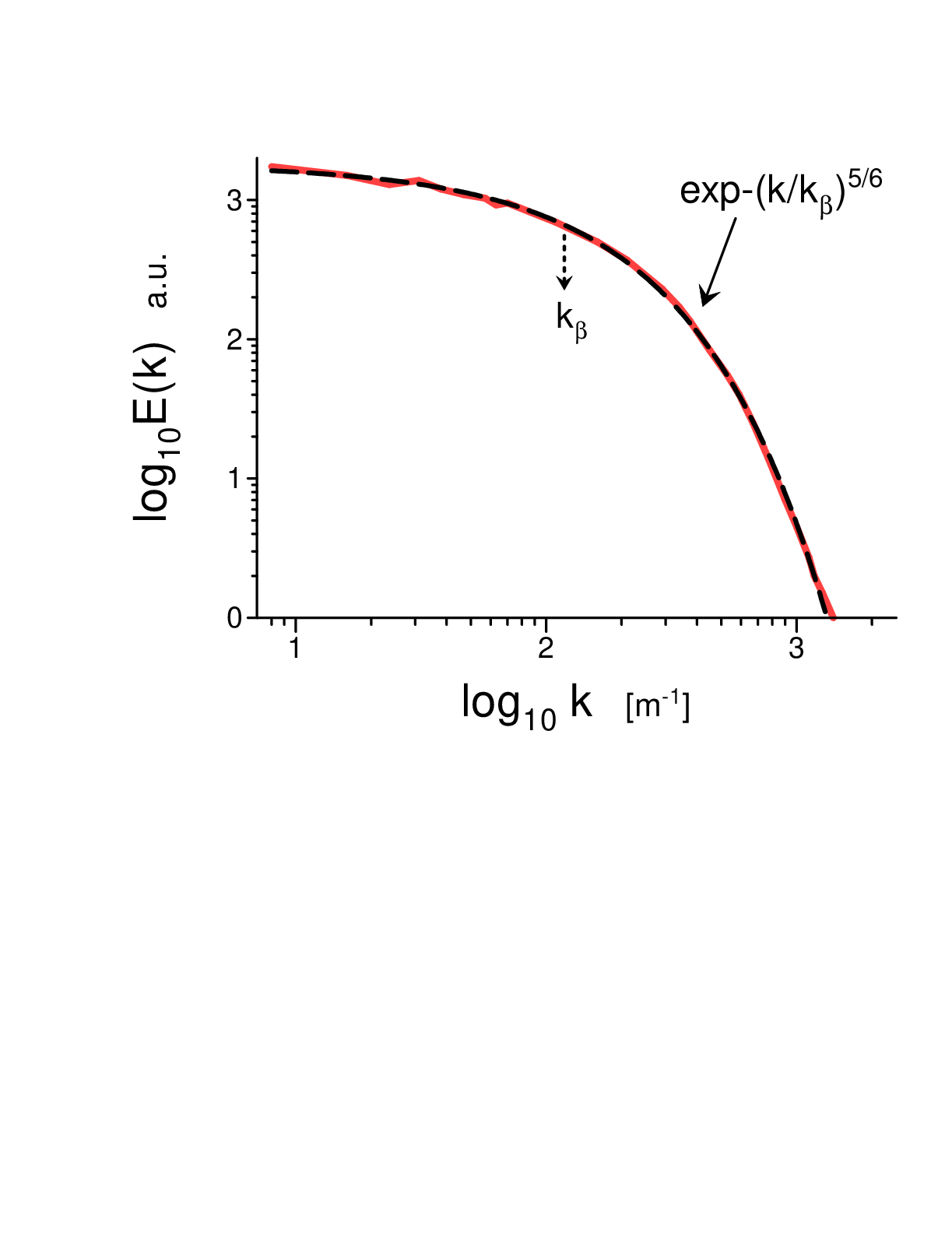} \vspace{-4.35cm}
\caption{Spectrum for the velocity fluctuations experimentally measured at the location behind the grid where the aerosol particles, droplets, and the chaotic fluid flow actively interacted. } 
\end{figure}

 The term ${\bf F}_p$ in the Navier-Stokes equation (11) is
$$
 {\bf F}_p = \frac{1}{m_f} \sum^{N}_{n=1} {\bf f}_n \eqno{(12)}
 $$
 here ${\bf f}_n$ is the drag force acting on particle $n$ and $m_f$ is the fluid mass within the control volume. \\

   The particle equation of motion was taken in the form 
$$
m_p\frac {d{\bf v}_p}{dt_p} = m_p\frac{{\bf u} - {\bf v_p}}{\tau_p}  +(m_p-m_f) {\bf g} \eqno{(13)} 
$$
 here ${\bf v}_p$ and ${\bf u}$ are the instantaneous velocities of the particle and of the carrier flow at
the particle location, $d/dt_p$ is the time derivative following the moving particle, $m_p$ is the particle mass, $\tau_p$ is the particle response time, ${\bf g}$ is the gravitational
acceleration. This equation is expected to be a good approximation for heavy particles ($\rho_p/\rho \gg 1$).\\

  The chaotic motion of the carrier flow was generated by initial conditions in the form of a random
Gaussian noise with an energy spectrum
$$
E_0(k) \propto k \exp-(k/k _0)   \eqno{(14)}
$$
The injection of the particles was done at $t=1$ in the DNS time terms. \\

  For the spectral data shown in Fig. 1 the `microparticles' correspond to the case of small Stokes number $St_{\eta} = \tau_p/\tau_{\eta} \ll 1$ (the $\tau_{\eta}$ is the Kolmogorov time scale), the `critical' particles correspond to $St_{\eta} \sim 1$, and the `large' particles correspond to large Stokes number $St_{\eta} \gg 1$ (at fixed particle volume fraction $\phi_v = 10^{-3}$ and particle mass loading $\phi_m =1$),  the gravity is not taken into account. For the spectral data shown in Fig. 2 the Stokes number was fixed $St_{\eta} =1$ and the particles volume fraction $\phi_v$ was varying. \\

  It was observed in the Ref. \cite{fe} that the microparticles were not centrifugally ejected from the vortical structures of the carrier fluid due to their fast response to the fluid velocity fluctuations. Therefore, the conditions of quasi-isotropy and quasi-homogeneity can be still applied to this case. The critical particles were ejected from the vorticity cores but were not completely ejected from the original vortical structures due to the centrifugal effect as the large particles do. The spectral consequences of these differences can be seen in Figs. 1 and 2 (cf Eqs. (9), (10), and (20)).\\
  
    Let us now discuss the results of a recent laboratory experiment reported in Ref. \cite{schum}. In this experiment, a moist-air wind tunnel flow behind a passive grid with the aerosol particles injected in the measuring section was used in order to experimentally simulate `aerosol–cloud–turbulence' interactions. Since the $Re_{\lambda} =30$ the flow was not turbulent but chaotic. This value of $Re_{\lambda}$ is very close to that corresponding to Fig. 1 (see above). Therefore, one can compare the results of this experiment with those shown in the Fig. 1, especially because the flows behind grids are usually used to simulate isotropic and homogeneous situations (the droplets in the flow were small). \\ 
    
    Figure 3 shows the spectrum for the velocity fluctuations measured at the location behind the grid where the aerosol particles, droplets, and the chaotic fluid flow actively interact. The dashed curve indicates the stretched exponential spectrum Eq. (10) (cf Fig. 1) and the dotted arrow indicates the position of $k_{\beta}$. \\

\section{Spontaneous breaking of local reflectional symmetry}

 The previous consideration was concentrated on isotropic, homogeneous flows with global (net) reflectional symmetry. The global reflection symmetry results in zero global (mean) helicity. The point-wise helicity, however, may not identically be equal to zero in this case. Spontaneous breaking of the local reflectional symmetry (and related spontaneous helicity fluctuations) can be considered as an intrinsic property of the chaotic/turbulent flows (see, for instance, Refs. \cite{bkt},\cite{kerr},\cite{hk}). The emergence of the moving with the fluid well-defined vorticity blobs with non-zero blob's helicity can accompany this phenomenon \cite{moff1}-\cite{bt}. Eventually, the vorticity blobs with high relative helicity and low viscous dissipation can be a generic property of the chaotic/turbulent flows \cite{moff2}. The low viscous dissipation in the blobs can result in adiabatic invariance of their helicity \cite{moff1},\cite{mt}. \\
 
   The well-known phenomenon of inertial particle ejection from the areas of high vorticity (strengthened by the high relative helicity) can preserve this adiabatic invariance for the particle-laden flows. For the microparticle-laden flows, in which the particles are mainly moving with the fluid (see above), this adiabatic invariance should be also naturally preserved. \\

   Since the global/net helicity should be zero at the spontaneous breaking of local reflectional symmetry the localized positive and negative blobs's helicities should be canceled at the overall average.\\

   The helicity in a vorticity blob is
$$
H_j = \int_{V_j} h({\bf r},t) ~ d{\bf r}.  \eqno{(15)}
$$
where $h({\bf r},t) = {\bf u} \cdot {\boldsymbol \omega}$ is helicity distribution, $V_j$ is spatial volume of the j-blob, $\bf{u}$ is velocity field, and ${\boldsymbol \omega} ({\bf r},t)= \nabla \times {\bf u}  ({\bf r},t)$ is vorticity field.\\

   The moments of the helicity distribution  $h({\bf r},t) = {\bf u} \cdot {\boldsymbol \omega}$ can be then defined as \cite{lt} ,\cite{mt}
$$
{\rm I_n} = \lim_{V \rightarrow  \infty} \frac{1}{V} \sum_j H_{j}^n  \eqno{(16)}
$$
here $V$ is the total volume of the blobs.\\

   Let $H_j^{-}$ be the helicity of the blobs with negative helicity and $H_j^{+}$ be the helicity of the blobs with positive helicity. Then corresponding sign-defined moments are
$$
{\rm I_n^{\pm}} = \lim_{V \rightarrow  \infty} \frac{1}{V} \sum_j [H_{j}^{\pm}]^n  \eqno{(17)}
$$ 
The summation in Eq. (17) is made over the blobs with negative (or positive) $H_{j}^{\pm}$ only.  \\

   Due to the global/net reflectional symmetry ${\rm I_n} = {\rm I_n^{+}} + {\rm I_n^{-}} =0$ for the odd $n$. Then for the odd $n$: ${\rm I_n^{+}} = - {\rm I_n^{-}}$.\\
 
  It should be noted that the blobs with high relative helicity provide the main contribution to the moments with high $n$ \cite{bt} (though, often the value $n=2$ can be already considered as sufficiently high for strongly intermittent chaotic/turbulent flows). This gives additional support for the use of the helicity moments ${\rm I_n}$ and ${\rm I_n^{\pm}}$ as relevant adiabatic invariants for the particle-laden flows.\\

\section{Distributed chaos and the helicity moments}

   Let us begin, for simplicity, with the flows dominated by the third moment of the helicity distribution ${\rm I}_3^{\pm}$ as a quasi-invariant. 

   If one uses the estimate Eq. (7), then from the dimensional considerations
$$
 u_c \propto |I_3^{\pm}|^{1/6}~ k_c^{1/2}    \eqno{(18)}
$$       
 and for the normal distribution of $u_c$  \cite{my} one obtains $P(k_c)$  
$$
P(k_c) \propto k_c^{-1/2} \exp-(k_c/4k_{\beta})  \eqno{(19)}
$$
  Substitution of the Eq. (19) into Eq. (3) gives 
 $$
 E(k) \propto \exp-(k/k_{\beta})^{1/2}  \eqno{(20)}
 $$ 
  
    One can compare Eq. (20) with the Figs. 1 and 2. \\

    In a general case of a  flow dominated by the quasi-invariant $I_n$ for an even moment or by the quasi-invariant $I_n^{\pm}$ for an odd moment one can write using the dimensional considerations
 $$
u_c \propto  |I_n^{\pm}|^{1/2n}~ k_c^{\alpha_n}   \eqno{(21)}
 $$  
 for the odd moments and 
 $$
 u_c \propto  I_n^{1/2n}~ k_c^{\alpha_n},    \eqno{(22)}
 $$  
 for the even moments.   Here
$$
\alpha_n = 1-\frac{3}{2n},  \eqno{(23)}
$$

  Then from the Eqs. (8) and (23) one obtains 
 $$
 \beta_n = \frac{2n-3}{3n-3}   \eqno{(24)}  
 $$  
   
 Let us consider two end cases: 
 
for $n \gg 1$
$$
E(k) \propto \exp-(k/k_{\beta})^{2/3},  \eqno{(25)}
$$ 

and for $n=2$ (i.e. for the Levich-Tsinober invariant \cite{lt})
$$
E(k) \propto \exp-(k/k_{\beta})^{1/3}  \eqno{(26)}
$$   

   Figure 4  shows the three-dimensional kinetic energy spectra computed in freely decaying homogeneous fluid motion for $Re_{\lambda} = 130$. The spectral data were taken from Fig. 3b of the Ref. \cite{dej}. In this DNS the gravitational acceleration ${\bf g}$ in the Eq. (13) was taken into account and the particle-laden flow was anisotropic. The Stokes number $St_{\eta} = \tau_p/\tau_{\eta}$ is the varying parameter, $\phi_v = 3\times 10^{-5}$, and $\rho_p/\rho =5000$.  \\
   
\begin{figure} \vspace{-0.8cm}\centering
\epsfig{width=.46\textwidth,file=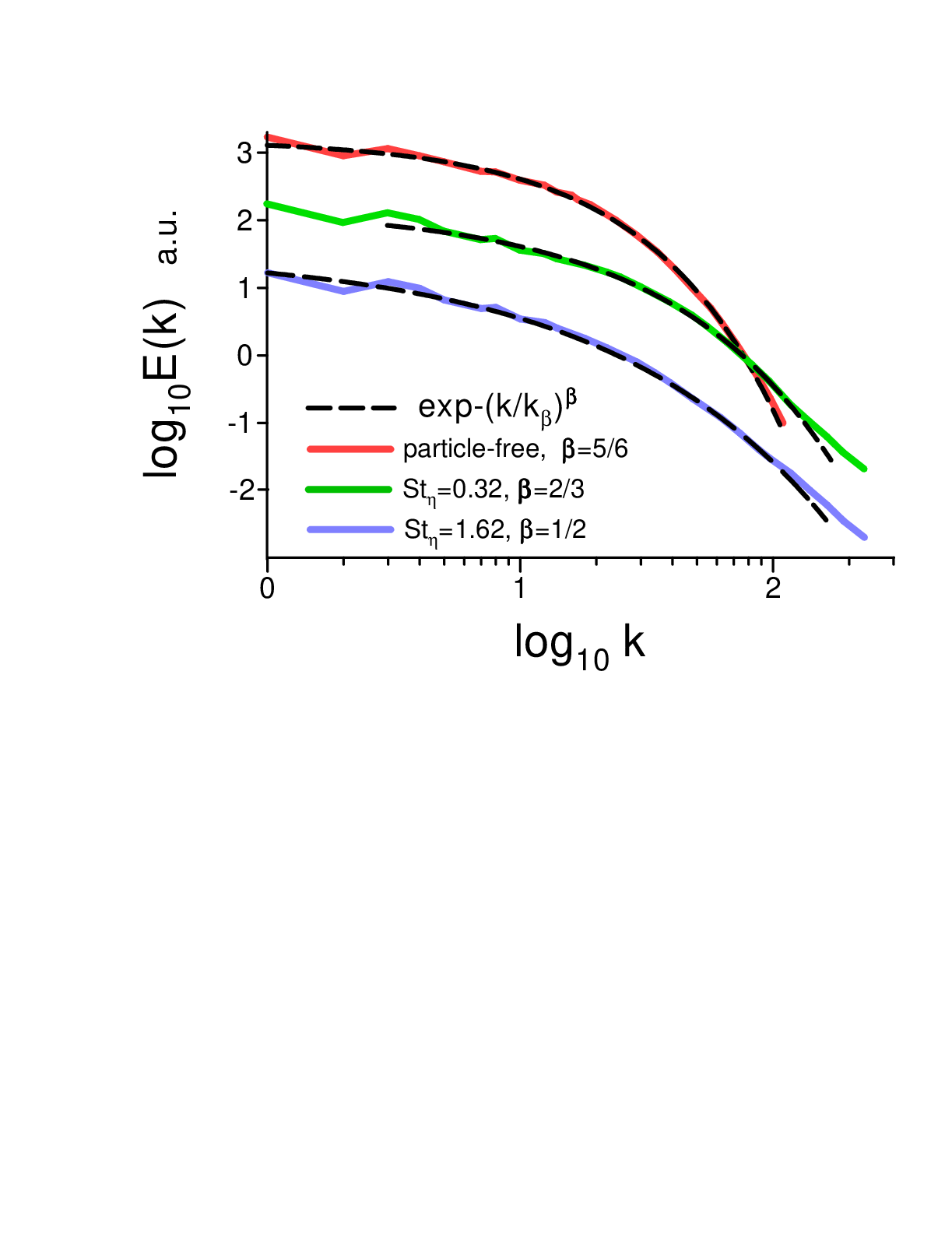} \vspace{-4.95cm}
\caption{Three-dimensional kinetic energy spectra computed in the homogeneous free decay at $t = 3$ for different values of the Stokes number $St_{\eta}$. The spectra are vertically shifted for clarity. } 
\end{figure}
   
\begin{figure} \vspace{-0.5cm}\centering\hspace{-1.2cm}
\epsfig{width=.48\textwidth,file=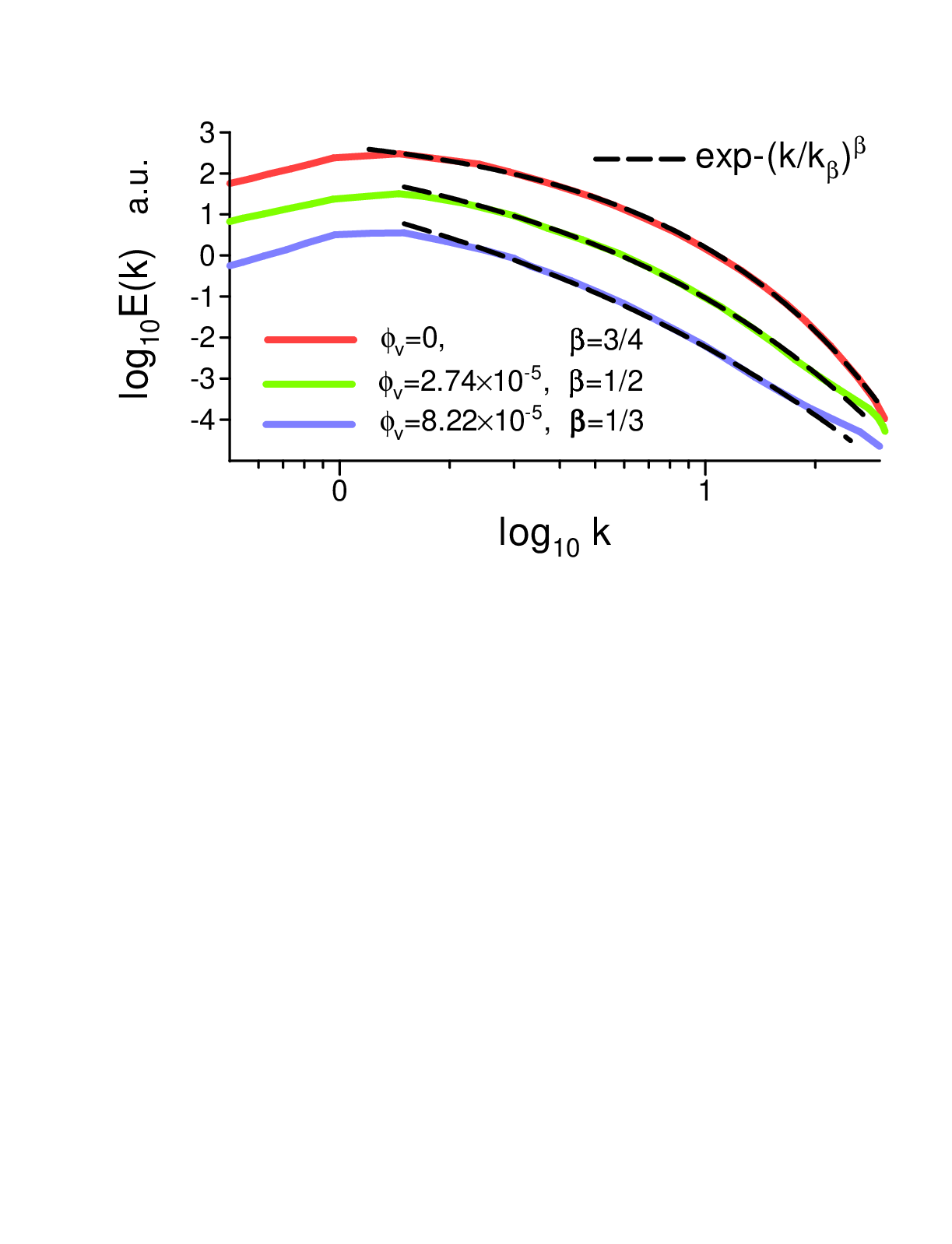} \vspace{-6.2cm}
\caption{ Kinetic energy spectra computed in a statistically {\it stationary} isotropic homogeneous fluid motion for $Re_{\lambda} = 35.4$, $St_{\eta} = 5$  (the spectra are vertically shifted for clarity). }
\end{figure}
\begin{figure} \vspace{-0.65cm}\centering \hspace{-1cm}
\epsfig{width=.48\textwidth,file=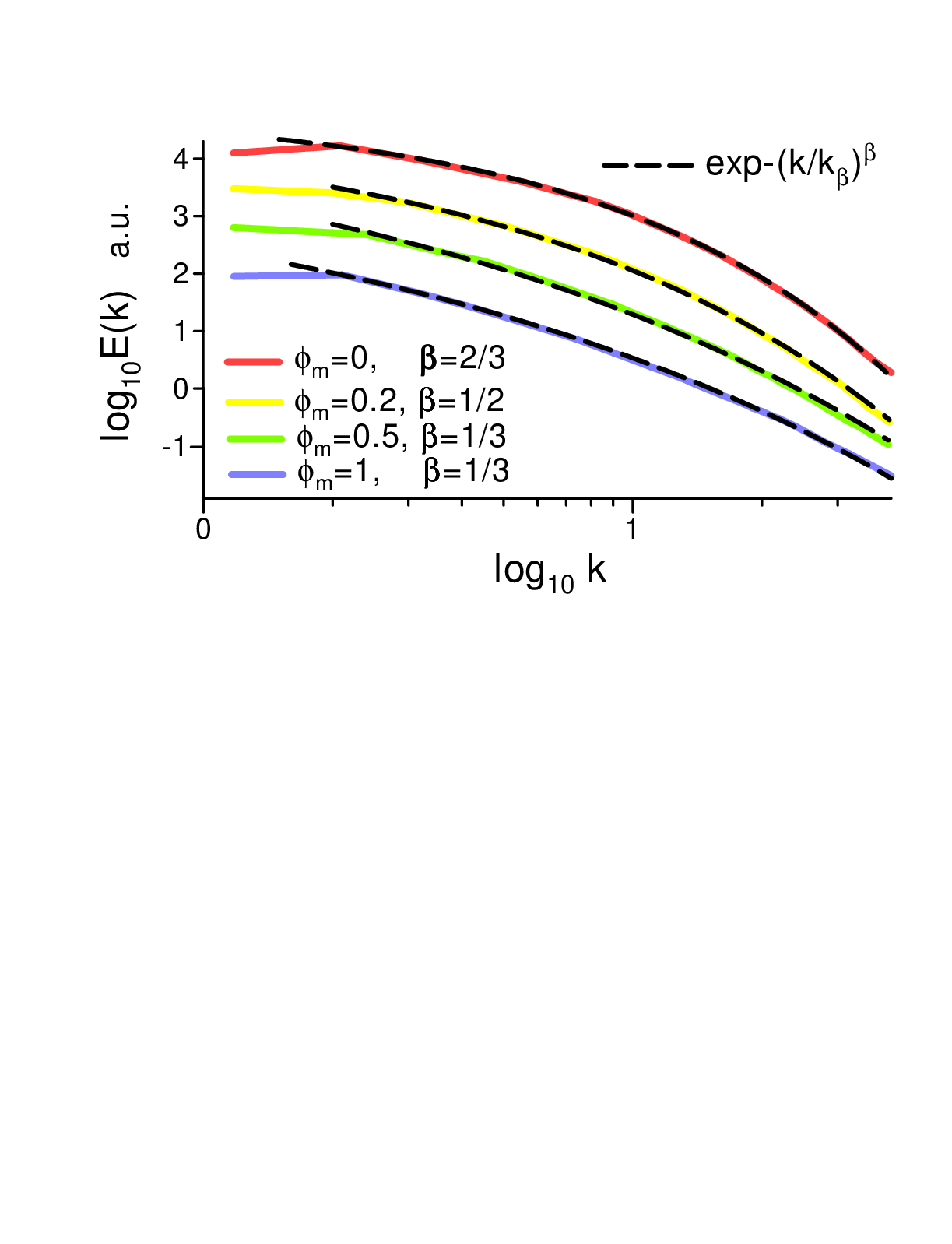} \vspace{-5.94cm}
\caption{Kinetic energy spectra computed in a statistically stationary isotropic homogeneous fluid motion for $Re_{\lambda} = 62$ (the spectra are vertically shifted for clarity).}
\end{figure}
\begin{figure} \vspace{-0.5cm}\centering \hspace{-1cm}
\epsfig{width=.47\textwidth,file=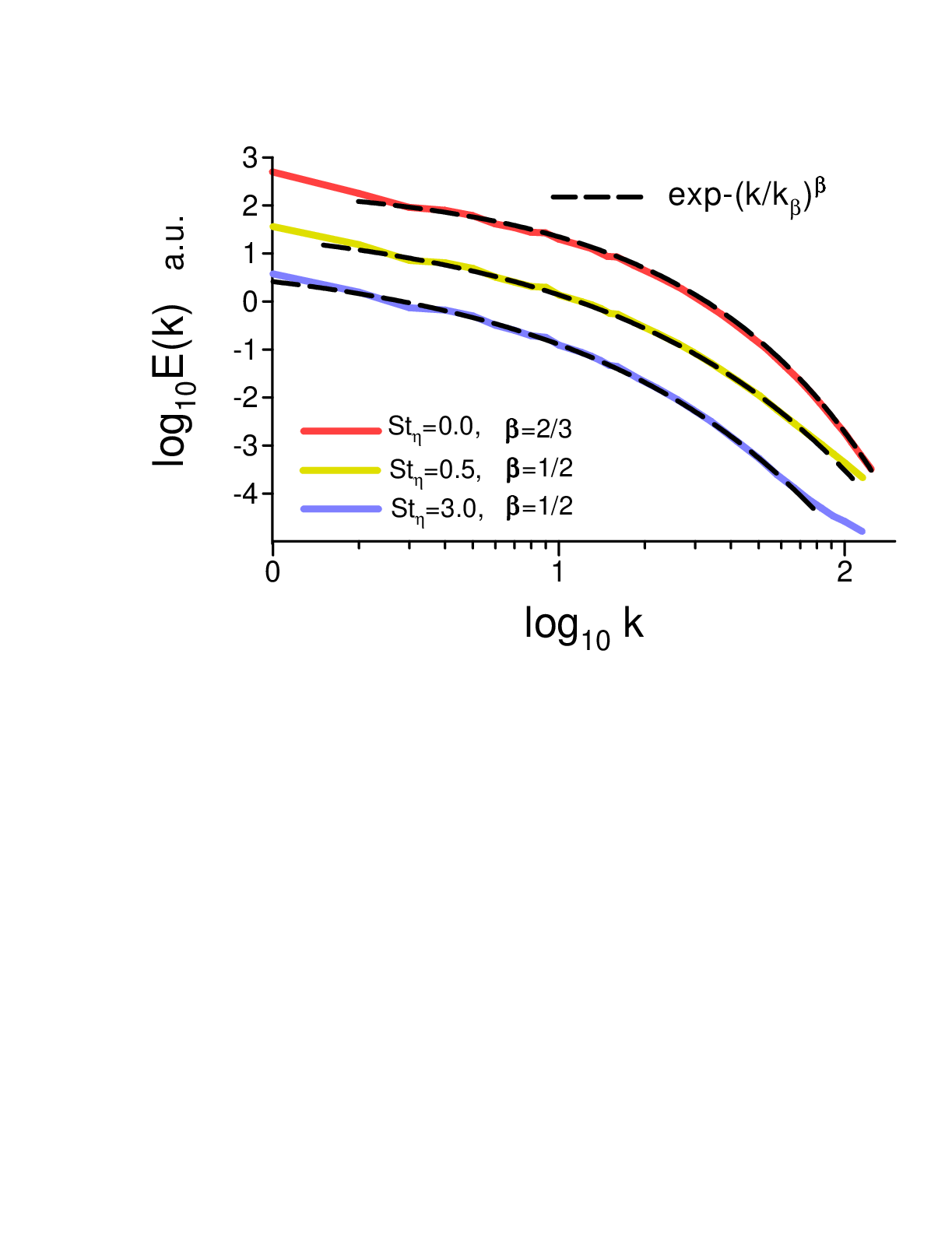} \vspace{-5.3cm}
\caption{Kinetic energy spectra computed in a statistically stationary isotropic homogeneous fluid motion for $Re_{\lambda} = 140$ (the spectra are vertically shifted for clarity).} 
\end{figure}

   The dashed curves indicate the stretched exponentials Eq. (10) (for the particle-free case), and Eqs. (25) and (20) (for the particle-laden cases). As for the previously considered cases of the isotropic free decaying situation without gravitation, the situation with gravitation exhibits the trend of the randomization enhanced by the particles. Unlike the free decaying cases with low $Re_{\lambda}$ (shown in the Figs. 1 and 2) in the case shown in the Fig. 4 the original particle-free flow was already in the state of distributed chaos with the stretched exponential spectrum Eq. (10) because the increase of $Re_{\lambda}$ results in enchanted randomization on its own \cite{b1}). \\
   
 Figure 5 shows the kinetic energy spectra computed in statistically {\it stationary} isotropic homogeneous fluid motion for $Re_{\lambda} = 35.4$. The spectral data were taken from Fig. 9 of the Ref. \cite{mgw}. It should be noted that the spectra shown in Figs. 1-4 were computed for the free decaying chaotic flows, whereas the statistically stationary state was reached in the DNS reported in the Ref. \cite{mgw} using an external random forcing. For the particle-laden cases the Stokes number was fixed $St_{\eta} =5$, and the particle volume fraction $\phi_v$ was varying. \\ 
 
   The dashed curves indicate the stretched exponentials Eq. (9) (for the particle-free case), and Eqs. (20) and (26) (for the particle-laden cases). As for the free decaying situation, the statistically stationary situation exhibits the trend of the randomization enhanced by the particles. Unlike the free decaying cases (shown in the Figs. 1 and 2) in the randomly forced case shown in the Fig. 5 the original particle-free flow was already in the state of distributed chaos with the stretched exponential spectrum Eq. (9). Therefore, one can observe in the Fig. 5 the spectrum Eq. (26) corresponding to the highly randomized distributed chaos. \\

   Figure 6 shows the kinetic energy spectra computed in a statistically stationary isotropic homogeneous fluid motion for $Re_{\lambda} = 62$. The spectral data were taken from Fig. 5a of the Ref. \cite{bss}. The particle mass loadings $\phi_m$ is the varying parameter in this case. The dashed curves indicate the stretched exponentials Eq. (25) (for the particle-free case), and Eqs. (20) and (26) (for the particle-laden cases). And again in the randomly forced case shown in the Fig. 6 the original particle-free flow was already in the state of distributed chaos with the stretched exponential spectrum Eq. (25) (there was an increase in $Re_{\lambda}$ compared with the previous example). Therefore, one can observe in the Fig. 6 the spectrum Eq. (26) corresponding to the highly randomized distributed chaos. \\
   
\begin{figure} \vspace{-0.6cm}\centering 
\epsfig{width=.46\textwidth,file=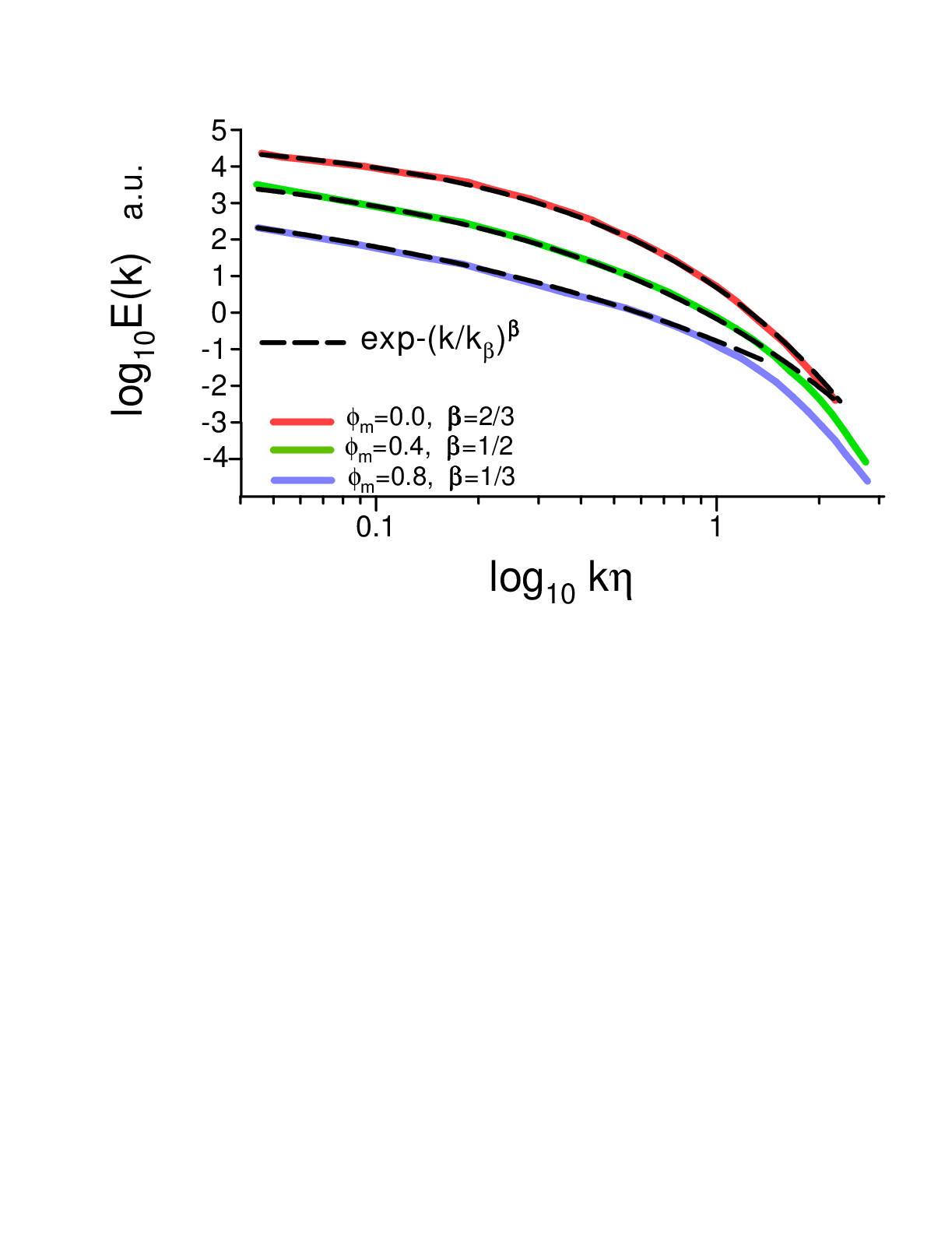} \vspace{-5.54cm}
\caption{Kinetic energy spectra computed in a statistically stationary {\it shear} homogeneous flow for $Re_{\lambda} = 80$ (the spectra are vertically shifted for clarity).} 
\end{figure}
   
   Figure 7 shows the kinetic energy spectra computed in a statistically stationary isotropic homogeneous fluid motion for $Re_{\lambda} = 140$. The spectral data were taken from Fig. 2 of the Ref. \cite{ci}. The Stokes number $St_{\eta}$ is the varying parameter (particle volume fraction $\phi_v = 2\times 10^{-4}$, the particle/fluid density ratio is equal to 1000). The dashed curves indicate the stretched exponentials Eq. (25) for the particle-free case and Eq. (20) for the particle-laden cases.\\ 
  
   Finally, let us consider the results of the numerical simulations of a particle-laden {\it shear} flow reported in a recent paper Ref. \cite{bat}. A new Exact Regularized Point Particle method was applied in the Ref. \cite{bat} to a homogeneous shear flow with the shear strength $S^{\star} =7$. \\
   
   Figure 8 shows the kinetic energy spectrum computed at $R_{\lambda} = 80$, $St_{\eta} = 1$ and different values of the mass loading parameter $\phi_m =$ 0, 0.4, and 0.8. The spectral data were taken from Fig. 4 of the Ref. \cite{bat}. The dashed curves indicate the stretched exponentials Eq. (25) for the particle-free case, and Eqs. (20) and (26) for the particle-laden cases with $\phi_m = 0.4$ and $\phi_m = 0.8$ respectively.
   
 \section{Conclusions}
   
  As one can see from the above-given examples an increase in the particle volume fraction, particle mass loading, and Stokes number results generally in stronger randomization of the particle-laden flows. The interplay of these parameters provides a rather complex picture (value of $Re_{\lambda}$ and presence or absence of the external random force also should be taken into account). \\
  
  Relevant dynamical invariants (both dissipative and non dissipative) dominate the randomization process and the spontaneous breaking of local reflectional symmetry plays an important role in this process providing a higher level of the randomization. \\ 
  
   The notion of distributed chaos can be instrumental in quantifying this phenomenon and the value of the parameter ${\beta}$ can be used as a measure of the randomization.
  
\section{Acknowledgments}

I thank H.K. Moffatt and J. Schumacher for useful information related to their papers.


\begin{thebibliography}{99}
\bibitem{fm} U. Frisch and R. Morf, Phys. Rev., {\bf 23}, 2673 (1981)
\bibitem{oh} N. Ohtomo, K. Tokiwano, Y. Tanaka et. al., J. Phys. Soc.
Jpn., {\bf 64}, 1104 (1995)
\bibitem{mm1} J. E. Maggs and G. J. Morales, Phys. Rev. Lett., {\bf 107},185003 (2011) 
\bibitem{mm2} J. E. Maggs and G. J. Morales, Phys. Rev. E {\bf 86}, 015401(R) (2012)
\bibitem{kds} S. Khurshid, D.A. Donzis and K.R. Sreenivasan, Phys. Rev. Fluids, {\bf 3}, 082601(R) (2018)
\bibitem{fe} A. Ferrante and S. Elghobashi, Phys. Fluids, {\bf 15}, 315 (2003).
\bibitem{et} S. Elghobashi and G.C. Truesdell, Phys. Fluids, {\bf 5}, 1790 (1993)
\bibitem{bir} G. Birkhoff, Commun. Pure Appl. Math., {\bf 7}, 19 (1954)
\bibitem{saf} P. G. Saffman, J. Fluid. Mech., {\bf 27}, 551 (1967)
\bibitem{dav} P.A. Davidson, J. Fluid Mech., {\bf 663}, 268 (2010)
\bibitem{my} A. S. Monin, A. M. Yaglom, Statistical Fluid Mechanics, Vol. II: Mechanics of Turbulence (Dover Pub. NY, 2007)
\bibitem{jon} D.C. Johnston, Phys. Rev. B, {\bf 74}, 184430 (2006)
\bibitem{schum} D. Niedermeier, J. Voigtlander1, S. Schmalfuß, D. Busch, J. Schumacher, R.A. Shaw, and F. Stratmann, Atmos. Meas. Tech., {\bf 13}, 2015 (2020)
\bibitem{bkt} A. Bershadskii, E. Kit, A. Tsinober, Proc. R. Soc. Lond. A, {\bf 441}, 147 (1993)
\bibitem{kerr} R.M. Kerr,  In: Elementary Vortices and Coherent Structures, Proceedings of the IUTAM Symposium Kyoto, 1-8 (2004)
\bibitem{hk} D.D. Holm, R.M. Kerr, Physics of Fluids, {\bf 19}, 025101 (2007)
\bibitem{moff1} H.K. Moffatt, J . Fluid Mech., {\bf 35}, 117 (1969)
\bibitem{moff2} H.K. Moffatt, J . Fluid Mech., {\bf 159}, 359 (1985)
\bibitem{lt} E. Levich and A. Tsinober, Phys. Lett. A {\bf 93}, 293 (1983)
\bibitem{mt} H.K. Moffatt and A. Tsinober, Annu. Rev. Fluid Mech., {\bf 24}, 281 (1992)
\bibitem{bt} A. Bershadskii and A. Tsinober,  Phys. Rev. E, {\bf 48}, 282 (1993)
\bibitem{dej} A. Dejoan, J. Phys.: Conference Series, {\bf 333}, 012006 (2011)
\bibitem{b1} A. Bershadskii, arXiv:2305.05554 (2023)
\bibitem{mgw} G. Mallouppas, W.K. George, and B.G.M. van Wachem, Int. J. Heat Fluid Flow, {\bf 67}, 74 (2017)
\bibitem{bss}  M. Boivin, O. Simonin, and K.D. Squires, J. Fluid Mech. {\bf 375}, 235 (1998)
\bibitem{ci}  M. Carbone and M. Iovieno, WIT Trans. Eng. Sci., {\bf 120}, 237 (2018)
\bibitem{bat} F. Battista, P. Gualtieri , J.-P. Mollicone, and  C.M. Casciola, Int. J. Multiphase Flow, {\bf 101}, 113 (2018)
\end{thebibliography}
\end{document}